# Yeild, Area and Energy Optimization in STT-MRAMs Using Failure-Aware ECC

ZOHA PAJOUHI, XUANYAO FONG, ANAND RAGHUNATHAN,
KAUSHIK ROY, Purdue University

Spin-Transfer Torque MRAMs are attractive due to their non-volatility, high density and zero leakage. However, STT-MRAMs suffer from poor reliability due to shared read and write paths. Additionally, conflicting requirements for data retention and write-ability (both related to the energy barrier height of the storage device) makes design more challenging. Furthermore, the energy barrier height depends on the geometry of the storage. Any variations in the geometry of the storage device lead to variations in the energy barrier height. In order to address poor reliability of STT-MRAMs, usage of Error Correcting Codes (ECC) has been proposed. Unlike traditional CMOS memory technologies, ECC is expected to correct both soft and hard errors in STT-MRAMs. To achieve acceptable yield with low write power, stronger ECC is required, resulting in increased number of encoded bits and degraded memory capacity. In this paper, we propose Failure-aware ECC (FaECC), which masks permanent faults while maintaining the same correction capability for soft errors without increased number of encoded bits. Furthermore, we investigate the impact of process variations on run-time reliability of STT-MRAMs. In order to analyze the effectiveness of our methodology, we developed a cross-layer simulation framework that consists of device, circuit and array level analysis of STT-MRAM memory arrays. Our results show that using FaECC relaxes the requirements on the energy barrier height, which reduces the write energy and results in smaller access transistor size and memory array area.

Categories and Subject Descriptors: Spintronics and magnetic technologies

General Terms: Non-volatile memory, STT-MRAMs, ECC, Erasure, Resistive memory, Yield, Run-time reliability,

## 1. INTRODUCTION

Spin-Transfer Torque (STT) memories are considered to be promising for future on-chip memory technology due to their favorable characteristics such as high density, non-volatility and near-zero leakage [Slonczewski, 1996; Berger, 1996; Katine et al. 2000; Li et al. 2008]. Nevertheless, they suffer from poor reliability that manifests in the form of low manufacturing yield, as well as run-time errors. Furthermore, ensuring high reliability through design leads to increased read and write energy and reduced density [Wu et al., 2009; Wang et al., 2013; Zhou et al., 2009; Xu et al., 2009; Del Bel et al., 2014; Kang et al., 2013; Yang et al., 2012].

Several research efforts have been devoted to addressing the poor reliability of STT-MRAMs at the device, circuit, and architecture levels [Pajouhi et al., 2015; Kwon et al., 2015; Kang et al, 2015; Wang et al., 2008; Apalkov et al., 2006]. Single-ended current sensing scheme utilized in STT-MRAM results in poor read reliability due to process variations in the electrical characteristics of the storage device (i.e., Tunneling Magneto-Resistance, TMR, and Resistance-Area, RA, product) that makes it difficult to distinguish between "1"s and "0"s reliably. On the other hand, since STT switching is a stochastic process [Kim et al., 2012; Fong et al., 2012], increased write currents are necessary to ensure reliable write operation. Reducing write current improves energy efficiency but increases the probability of write errors and results in degraded yield. In order to reduce write errors, the Energy Barrier ($E_B$) height of the MTJ can be reduced [Li et al., 2008; Augustine et al., 2010]. However, the retention

This research was funded in part by the Center for Spintronics: Materials, Interfaces and Architecture, a StarNet Center funded by DARPA and MARCO, by Semiconductor Research Corporation, and by National Science Foundation.

time of the MTJ depends on the energy barrier height [Naemi et al., 2013], which needs to be high enough to ensure sufficiently long retention time.

From the above discussion, it is evident that there is a trade-off between the reliability and energy consumption of the memory array. Furthermore, high yield and run-time reliability impose conflicting requirements on the energy barrier height. This raises the question of whether it is possible to address reliability concerns in STT-MRAMs without foregoing their benefits such as high density and non-volatility. One of the most prevalent methods to address reliability in memories is the use of Error Correcting Codes (ECC). However, ECC imposes overheads in the form of increased area and power consumption for the additional encoding bits and encoder/decoder circuitry. On the other hand, it enables more efficient read and write operations in STT-MRAM bit-cells, and can result in enhanced yield. Recent research has shown that ECC can be used to improve the density and energy efficiency of STT-MRAMs [Xu et al, 2009; Del Bel et al., 2014; Pajouhi et al. 2015; Kwon et al. 2015]. These research efforts focus on enhancing reliability; however, they do not distinguish between yield and run-time reliability. In [Del Bel et al., 2014] the authors consider ECC and address run-time reliability. However, they do not consider the impact of process variations on the run-time reliability and retention time. Note, the above results suggest that in order to meet an acceptable yield and to maintain run-time reliability simultaneously, ECC with increased correction capability is required. However, as the correction capability is increased, the advantages associated with ECC insertion deteriorates due to increased overheads.

In this paper, we investigate the STT-MRAM bit-cell and explore the impact of process variations on different failure mechanisms. Specifically, we explain the impact of process variations on the run-time reliability and retention time of the memory array and analyze the impact of retention time on write power dissipation and write failures. In order to enhance the reliability of the memory array, we propose Failure-aware ECC (FaECC), to mask the permanent faults while maintaining the correction capability for soft errors. In FaECC scheme, we identify permanent defective bit-cells and exploit the knowledge of the location of the defective bit-cell within the encoded word to enhance the correction capability of ECC. In order to analyze the impact of FaECC, we developed a cross-layer simulation framework at the device, circuit and array levels of design abstraction. This framework is used to analyze the impact of different reliability enhancement techniques on the yield and run-time reliability of the memory array. In summary, we make the following key contributions:

- We provide a comprehensive discussion and statistical analysis of the relationship between parameters of the storage device and retention failures in STT-MRAMs in the presence of process variations.
- We develop a cross-layer simulation framework to evaluate the impact of process variations on the yield and run-time reliability of an STT-MRAM array. We utilize the simulation framework to analyze the impact of ECC on yield and run-time reliability. We show that using ECC to improve yield has a negative impact on the ability of ECC to improve run-time reliability.
- We propose a Failure-aware ECC (FaECC) to mask permanent faults without compromising the correction capability for transient faults. With such an approach, Single Error Correction and Double Error Detection (SECDED) is employed to correct transient faults as well as masking permanent faults simultaneously. Furthermore, by using this approach, the required energy barrier height of the memory array to meet the required run-time reliability decreases. This decrease results in reduced access transistor size and reduced read/write power.

The rest of the paper is organized as follows. In Section 2, we explain STT-MRAM preliminaries and describe different bit-cell failure mechanisms. In Section 3, we discuss the run-time reliability of STT-MRAM array. In Section 4 we propose FaECC, a Failure-aware ECC scheme for STT-MRAMs that exploits an understanding of failure mechanisms to improve yield without sacrificing run-time reliability. In Section 5 we present the cross-layer simulation framework used to evaluate the proposed FaECC scheme. In Section 6 we discuss the results obtained from our simulation framework. Section 7 provides the concluding remarks.

## 2. STT-MRAM PRELIMINARIES

An STT-MRAM bit-cell consists of a storage and an access transistor, as shown in Fig. 1. The Magnetic Tunnel Junction (MTJ) is the storage device of STT-MRAM and the access transistor is used to access the MTJ. The MTJ consists of two ferromagnetic layers – a pinned layer and a free layer – sandwiching a tunneling oxide (typically MgO). The pinned layer has a fixed magnetization orientation, while the magnetization of the free layer can be changed. The relative magnetic orientation of the free and the fixed layers determines the data stored in the MTJ. If the magnetization orientation of the free layer is the same as the fixed layer, they are said to be in parallel; however, if they are in opposite directions, they are said to be anti-parallel (we assume logic "0" is represented by the parallel orientation and "1" by anti-parallel). The magnetization orientation can be aligned with the surface of the MTJ – in-plane magnetic anisotropy (IMA) – or it can be perpendicular to the surface (PMA).

In order to write into the bit-cell, the word line is activated and a bias voltage is applied between the bit line and the source line to pass current through the MTJ. The direction of current flow defines the data value that is written into the bit-cell. The amount of write current needed, called the critical switching current, depends on the desired write time. Achieving acceptable write latencies typically requires high switching current, negatively impacting energy efficiency and reliability [Fong et al.,2012].

In order to read the bit-cell, the word-line is enabled and a bias voltage is applied between the bit-line and the source line, causing a current to pass through the MTJ. The current is then sensed to evaluate the resistance of the MTJ and to distinguish between a logic "1" and a logic "0". The read current should be substantially lower than the critical switching current of the MTJ to avoid accidentally writing into the bit-cell during read operations.

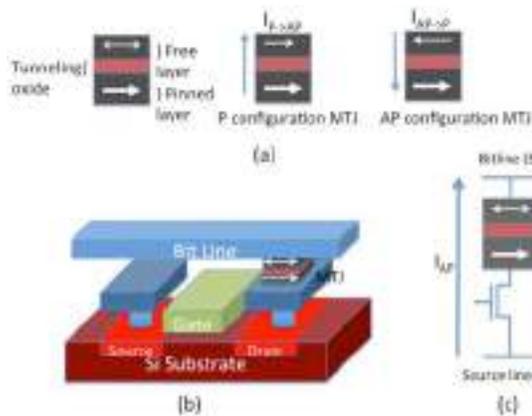

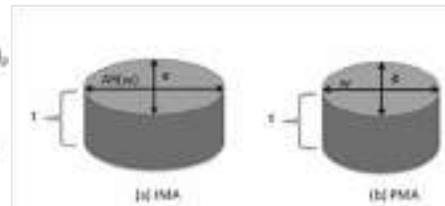

Fig. 1. (a) MTJ structure (b) bit-cell structure (c) standard connection.

Fig. 2. Free layer dimensions for (a) In- plane Magnetic anisotropy and (b) Perpendicular magnetic anisotropy.

There are four major failure mechanisms in STT-MRAMs: read decision failures, read disturb failures, write failures and retention failures [Fong et al.,2012].

Read decision failures occur due to an inability to correctly detect the value stored in the MTJ. As explained earlier, a voltage ($V_{read}$) is applied between the source line and the bit line and the data is determined by comparing the bit-cell current with a reference current ($I_{ref}$). Ideally, the bit-cells with different stored values have different currents passing through them (e.g., $I_P$ for the parallel configuration and $I_{AP}$ for anti-parallel) and the sensing margin is maximized by setting the reference current to the average of the two currents. However, due to process variations (e.g., variation in the RA product), the current passing through each bit-cell may differ from its nominal value. Therefore, $I_{ref}$ should be chosen carefully to minimize decision failures. Once $I_{ref}$ is defined, the sense amplifier is adjusted accordingly. Note that read-decision failures may be considered to be stuck-at-fault failures [Kang et al. 2015, Su et al. 2004].

Disturb failures occur when the data stored in the bit-cell is unintentionally overwritten during a read operation. This is due to increased current flowing through the MTJ during the read operation. Note that, since the direction of the read current matches that of only one of the write currents, this type of failure occurs only for one value of data (either "0" erroneously changing to "1" or vice versa). Disturb failures occur due to increased current drivability of the access transistor or decreased critical current of the MTJ. This decrease may be the result of process variations or thermal effects.

Write failures occur due to unsuccessful MTJ state change during write operation. They occur due to decreased current drivability of the access transistor or increased MTJ critical current due to process variations or thermal effects. Standard connection [Nebashi et al. 2009; Lin et al.,2009] of the bit-cell is considered in this work to mitigate write failures.

Finally, retention failures occur due to thermal effects. If thermal effects are large enough to flip the magnetization of the free layer, the MTJ changes its state. Retention failures are characterized by the retention life-time of the nanomagnet. The probability of retention failure in a single memory bit-cell at time $t$ is given by [Naemi et al., 2013]:

$$P_{FAIL\_THERMAL} = 1 - F_f(t), F_f(t) = \exp\left(\frac{-t}{t_{life}}\right), f_f(t) = \frac{1}{t_{life}}\exp\left(\frac{-t}{t_{life}}\right) \quad (1)$$

where $f_f(t)$ is the probability density function failure, $P_{FAIL\_THERMAL}$ is the probability of failure at time $t$ and $F_f(t)$ is the cumulative probability density function. Also, $t_{life}$, which is called the life-time of the MTJ, depends on the free layer characteristics [Augustine et al., 2010].

As observed in Eq. 1, the probability of retention failures depends on the time elapsed after the write event. Furthermore, the life-time of the bit-cell depends on the physical characteristics of the free layer. Note, due to process variations (leading to variations in the barrier height), some bit-cells are more vulnerable to retention failures as compared to the others.

In the next Section, we will investigate the impact of process variations on retention failures and its effect on run-time reliability.

## 3. RUN-TIME RELIABILITY ANALYSIS

### 3.1 Thermal Stability and Retention Time

Although STT-MRAMs are referred to as non-volatile memories, their ability to retain the stored data is limited in practice. As explained in the previous Section, the retention failure probability can be expressed in terms of the time elapsed from the time data was stored in the memory and the life-time of the free layer. The life-time of the free layer can in turn be expressed as [Augustine et al.,2010]:

$$t_{life} = (10^{-9})\exp\left(\frac{E_B}{K_B T}\right) \tag{2}$$

where $E_B$ is the Energy Barrier height, $K_B$ is the Boltzmann constant and $T$ is the temperature in Kelvin. $E_B$ depends on the geometric dimensions of the free layer. Fig. 2 illustrates the physical dimensions of the free layer for different magnetic anisotropy configurations. $E_B$ for an In-plane Magnetic Anisotropy (IMA) free layer can be expressed as [Apalkov et al. 2010]:

$$E_B = \frac{H_k M_S V}{2} \approx \frac{4\pi M_S t(AR-1)}{wAR} M_S \frac{\pi}{4} w^2 \cdot AR \cdot t \propto t^2 w(AR-1) \tag{3}$$

where $M_S$ is the saturation magnetization, $H_K$ is the effective field anisotropy, and $V$ is the volume of the free layer. Furthermore, $w$, $AR$ and $t$ are the width, aspect ratio and the thickness of the free layer respectively. As expressed in Eq. 3, $E_B$ depends on the geometry of the free layer and is therefore sensitive to process variations.

For a free layer with Perpendicular Magnetic Anisotropy, $E_B$ can be expressed as [Augustine et al., 2010]:

$$E_B = K_{u2}V = \frac{H_k^C M_S V}{2} = \frac{H_k^C M_S \frac{\pi}{4} w^2 t}{2} \propto t \, w^2 \tag{4}$$

where $K_{u2}$ is the uniaxial anisotropy, $V$ is the volume of the free layer and $H_k^C$ is the effective field anisotropy. As observed, $E_B$ also depends on the geometry of the free layer.

In order to ensure reliable operation of STT-MRAM, $E_B$ should be adjusted such that the requirements of run-time reliability are met. A typical memory reliability specification can be expressed in terms of FIT or failures in time, where 1 FIT is one failure per billion (devices × hours):

$$\frac{1}{\lambda} * (10^9) = 1 \, FIT \tag{5}$$

where $\lambda$ is the failure rate in hours and can be expressed as the equivalent of Mean Time To Failure (MTTF):

$$MTTF = \int_0^\infty t \, f_f(t) \, dt \tag{6}$$

where $f_f$ is the probability density function of time to failure, if and only if this integral exists (as an improper integral).

Therefore, for an MTJ device, we have:

$$MTTF = \left| \int_0^\infty \frac{t}{t_{life}} \exp\left(-t/t_{life}\right) dt \right| = t_{life} \tag{7}$$

If only a single device is considered, 1 FIT translates into 0.00876% failure over 10 years, and the required $E_B$ to meet the requirement of 1 FIT is about 50 $K_B T$. However, for larger memory arrays, 1 FIT should be considered for the entire memory array and not just a single MTJ device. For this purpose, let us consider a memory array with $n$ bit-cells. Under such conditions, the probability of correctness for the array can be defined as:

$$F_{farray} = \prod_{k=1}^n F_{fk} = \prod_{k=1}^n \exp\left(-t/t_{life}\right) = \exp\left(-nt/t_{life}\right) \tag{8}$$

In which $F_{farray}$ is the cumulative probability density function, therefore, the probability density function can be written as:

$$f_{farray} = (\frac{n}{t_{life}}) \exp(-\frac{nt}{t_{life}}) \tag{9}$$

the $MTTF$ for the memory array can be defined as follows:

$$MTTF_{array} = \left| \int_0^\infty \frac{nt}{t_{life}} \exp(-nt/t_{life}) \, dt \right| = \frac{t_{life}}{n} \tag{10}$$

In order to obtain the required $E_B$, the required $MTTF$ should be obtained from Eq. 6. Next, the life-time should be defined to meet the required $MTTF_{array}$ by solving Eq. 10 for the desired array size. Once the $E_B$ is derived, the free layer physical characteristics can be derived. In order to define the free layer characteristics, if the free layer is an IMA (PMA), Eq. 3 (4) should be used. In order to analyze run-time reliability, without confining to a set of MTJ parameters, the thermal stability factor is defined as follows:

$$E_{BN} = E_B/K_B T \, , \tag{11}$$

In the following Sections, we will derive the free layer physical characteristics based on the operating temperature and characteristics of the MTJ. Fig. 3 shows the required $E_{BN}$ for larger memory arrays (ignoring parameter variations) for 1 FIT. As observed, the required $E_{BN}$ increases with increasing memory size. Since the reliability metric of 1 FIT is kept constant, as the number of bit-cells in the memory array increases, the tolerable probability of failure for each bit-cell decreases. In order to meet this decreased probability, the $E_B$ should be increased.

### 3.2 The effect of ECC on run-time reliability

ECC is one of the most effective methods to improve reliability of memory arrays. Among different ECC codes, Bose-Chaudhuri-Hocquenhgem (BCH) codes are commonly used in memory arrays [Michelson et al.,1985]. A BCH code changes a $k$-bit word data into an $n$-bit word data by adding ($n$-$k$) bits to the word. The choice of $n$ depends on the desired correction capability of ECC. The choice of the word length at which ECC should be applied ($k$) and the extra bits ($n$-$k$) impacts the correction capability as well as the overheads incurred. The probability of correctness for an $n$-bit word with $m$-bit correction capability, with a bit error probability of $P_b$ can be expressed as:

$$P_{word} = \sum_{i=0}^m \binom{n}{i} (1-P_b)^{n-i} P_b^i \tag{12}$$

Furthermore, if the memory array has $s$ words, every word has to be encoded using the ECC scheme selected for the memory array. Then, the probability of correctness of the entire memory array would be:

$$P_{corr} = P_{word}^s \tag{13}$$

In order to obtain the required $E_{BN}$ for an array (with ECC), it is required to substitute the probability of failure of a single MTJ obtained in Eq. 1, into Eq. 12 as follows:

$$P_{word} = \sum_{i=0}^m \binom{n}{i} (\exp(-\frac{t}{t_{life}}))^{n-i} \left(\exp(-\frac{t}{t_{life}})\right)^{n-i} * \left(1 - \exp(-\frac{t}{t_{life}})\right)^i (1 - \exp(-t/t_{life}))^i \tag{14}$$

Next, the resultant $P_{word}$ is inserted into Eq. 13. At the next step, the probability density function is derived from the cumulative density function as follows:

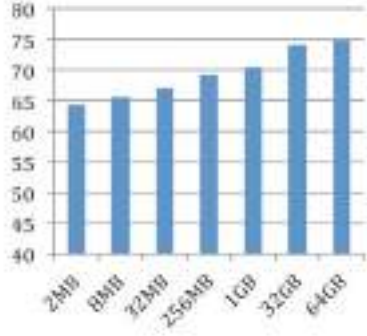

Fig. 3. Required $E_{BN}$ vs. memory size for 1 FIT.

$$f_{fcorr} = s \left( \frac{n}{t_{life}} \exp\left(-\frac{nt}{t_{life}}\right) + \sum_{i=1}^{m} \binom{n}{i} \exp\left(-\frac{(n-i)t}{t_{life}}\right) * \left(1 - \exp\left(-\frac{t}{t_{life}}\right)\right)^{i-1} * \frac{1}{t_{life}} \left[(n-i) - \right.\right.$$
$$\left.\left. n * \exp\left(-\frac{t}{t_{life}}\right)\right]\right) * \left(\sum_{i=0}^{m} \binom{n}{i} \exp\left(-\frac{(n-i)t}{t_{life}}\right) * (1 - \exp\left(-t/t_{life}\right))^{i}\right)^{s-1} \qquad (15)$$

Finally, the probability density function of time to failure is derived and inserted into Eq. 6 to obtain the $MTTF_{array}$:

$$MTTF_{array} = \left| \int_0^\infty t f_{fcorr} dt \right| \qquad (16)$$

For example, let us assume that the desired size of the memory is 4 MB and we apply ECC to every 128 bits in the array. Therefore, $k$=128 and $s$=4MB/128. For an ECC with Single Error Correction and Double Error Detection (SECDED) capability, the encoding should be performed $GF(2^s)$ where $GF(2^{deg})$ is the Galois field with degree $deg$. The number of additional bits is 8 for Hamming code (which is considered the simplest BCH code) and a single parity bit is added to detect an additional error, resulting in a total of 9 bits. Therefore, $m$=1 and $n$=9+128=137. These values should be inserted in Eq. 15 and Eq. 16 to obtain the $MTTF_{array}$. However, for a given $MTTF_{array}$, this process should be inverted to obtain the required $t_{life}$. Once the $t_{life}$ is derived, it can be inserted in Eq. 2 and 11 to obtain $E_{BN}$ and $E_B$. Fig. 4 illustrates the required $E_{BN}$ to meet the reliability level of 1 FIT for a 4 MB memory array with the word size of 128 bits. As observed, the required $E_{BN}$ decreases with an increase in the correction capability of ECC.

Further, let us consider that ECC is employed to correct read and write failures as well as retention failures. Under such conditions, the words that happen to contain such failures (read or write) and retention failures cannot be corrected with ECC. Moreover, if there are a large number of such words, the $E_B$ cannot be reduced as suggested above. In the next subsection, we will analyze the impact of such conditions on the efficacy of ECC for retention failures.

### 3.3 The impact of read and write failures on the efficacy of ECC

Let us consider a scenario where ECC is used for enhancing yield as well as run-time reliability. The presence of hard failures in a data word degrades its capability to correct retention, read and write failures. In order to determine the impact of degraded ECC capability on run-time reliability, let us assume that the number of words with $j$ non-retention failures is $n_j$. Then, the probability of correctness for the whole memory can be defined as:

$$P_{corr} = \prod_{j=0}^{m} P_{word j}^{n_j} \qquad (17)$$

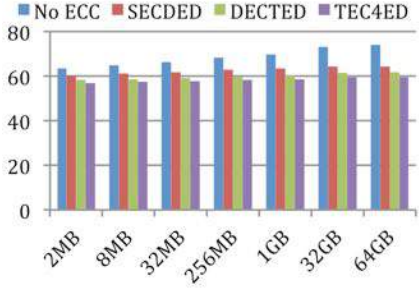

Fig. 4. Required $E_{BN}$ for different memory array sizes including ECC.

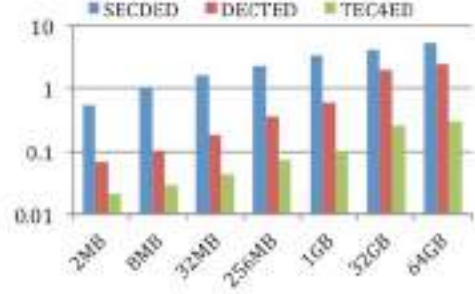

Fig. 5. Percentage of required increase in the $E_{BN}$ for an array with probability of defective bit-cell of 1e-5 if ECC is utilized for yield enhancement and improving run-time reliability.

where $m$ is the maximum number of correctable errors and $P_{wordj}$ can be obtained from the following equation:

$$P_{wordj} = \sum_{i=0}^{m-j} \binom{n}{i} (1 - P_b)^{n-i} p_b^i \qquad (18)$$

where $n$ is the total number of bits in a word. For example, let us consider a 4 MB array in which ECC with SECDED capability is used. Furthermore, let us assume that ECC is applied to a word length of 128 bits, which implies that the number of words in the array is $s=4MB/16B$. Additionally, let us assume that there are $b$ words with a single read or write failure and we have used ECC to correct these failures as well. If a retention failure also occurs in one of the $b$ lines mentioned above and in one of the bit-cells without read or write failures, ECC with SECDED capability will be unable to correct it because there is already a read or write failure in the same word. Note, in this example, $m=1$, $n_0=s-b$, $n_1=b$. In order to calculate the $MTTF_{array}$, the probability of correctness obtained in Eq. 1 should be substituted in Eq. 18. Following these steps, the probability of correctness for the memory array can be expressed as follows:

$$P_{corr} = \exp\left(-\frac{(ns-b)t}{t_{life}}\right)\left(n + (1-n)\exp\left(-\frac{t}{t_{life}}\right)\right)^b \qquad (19)$$

Next, the probability density function is derived similar to Eq. 15.

$$f_{corr} = \exp\left(-\frac{(ns-b)t}{t_{life}}\right)\left(n + (1-n)\exp\left(-\frac{t}{t_{life}}\right)\right)^{b-1}\left[\frac{ns(n+(1-n)\exp\left(-\frac{t}{t_{life}}\right))-bn}{t_{life}}\right] \qquad (20)$$

Finally, the required $E_{BN}$ and $E_B$ can be derived by substituting $f_{corr}$ into Eq. 6.

As an example, let us consider the same 4 MB memory array with 128-bit ECC word length. Furthermore, let us assume that the probability of having a read or write failure in each bit-cell is 1e-5 and that these failures are uniformly distributed among all the bit-cells. Next, based on the uniform distribution of these failures, we obtain the mean number of words with $j$ read or write failures ($n_j$) and insert the obtained $n_j$ into Eq. 19 and the probability density function should be derived similar to Eq. 20.

Fig. 5 illustrates the percentage increase in the required $E_{BN}$. As observed, the increase in the required $E_{BN}$ decreases with an increase in the correction capability of ECC. In order to avoid increasing $E_{BN}$, either an ECC scheme with higher correction capability is required or if possible, other yield enhancement techniques, such as redundant rows or columns can be introduced. However, due to poor reliability of STT-MRAM bit-cells, redundancy is not an effective method to enhance yield; it incurs high overheads [Kwon et al. 2015]. Therefore, a recent trend towards ensuring reliability is to utilize ECC for yield enhancement. However, in order to maintain the high yield and run-time reliability simultaneously, there is a need to use ECC with higher correction capability, requiring higher storage area. We propose a Failure-aware ECC (FaECC) scheme, which enhances the correction capability without adding large number of encoded bits to the array. In FaECC, we identify the read-decision failures, and use our proposed technique (described in the following section) to correct these failures. However, due to the stochastic nature of the write in STT-MRAMs, this method is not used to mitigate write failures.

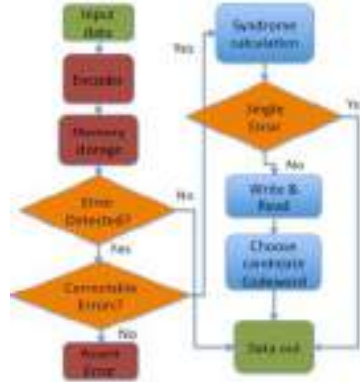

Fig. 6. FaECC procedure.

## 4. FAILURE-AWARE ECC BASED CORRECTION

Due to their simple structure and decoding scheme, BCH codes [Wilkerson et al., 2010; Strukov, 2006] are commonly used in memory design. Specifically, the Hamming code, which can be viewed as a special case of BCH codes, has found widespread use in memory ECC. The number of additional encoded bits required for ECC is determined by the desired correction capability and the word length at which ECC is applied.

Furthermore, ECC can be employed to correct errors with known locations. These types of errors are called erasures [Evain et al., 2014]. The concept of potentially-erroneous bits with recognized locations is well-known and is employed in digital communications but amazingly not commonly used in memory systems [Evain et al., 2014]. Theoretically, a code with minimum Hamming distance $d$ can correct $t$ random errors and $r$ erasures if $d>2*t+r$ [Walker et al., 1979; Carter et al., 1976; Siewiorek et al., 1998; Chen et al., 1984; Evain et al., 2014; N. H. Seong, 2010; R. E. Fujiwara, 1989]. Therefore, if we know all positions of errors, we can introduce $t=0$ and use the code for correcting erasures.

In FaECC, we use the concept of correcting erasures through ECC [Evain et al., 2014]. Fig. 6 illustrates the FaECC methodology. In this methodology, a SECDED code is used and erasure information are used to enable Double Error Correction (DEC) capability for stuck-at-fault errors. In this method, the encoding and decoding are performed similar to a SECDED encoding scheme and DEC decoding is enabled only if double errors are detected by the normal SECDED decoder. Once the DEC decoding is enabled, the erasure information are retrieved and used to correct the stuck-at-fault errors. In the next Subsection, we explain the FaECC scheme in detail.

## 4.1 Failure-Aware ECC scheme

In Hamming code, the data bits are encoded through an encoder to obtain an encoded word (c) to be stored in the memory:

$$c \equiv a.G \tag{21}$$

in which $a$ is the input word expressed as:

$$a = \{x_1, x_2, \ldots, x_k\}, x_i \epsilon \{0,1\} \tag{22}$$

Note that $k$ is the number of data bits. Further, $c$ is expressed as:

$$c = \{y_1, y_2, \ldots, y_n\}, y_i \epsilon \{0,1\} \tag{23}$$

in which $n$ is the total number of bits to be stored including encoded bits, the expression $\equiv$ is equivalent to:

$$c = (a.G) mod\ 2 \tag{24}$$

The encoded word is stored in the memory. Once the code-word is read from the memory ($\tilde{c}$), it may contain one or more errors. In regular Hamming decoders, the syndrome, $z$, can be calculated as follows:

$$z \equiv \tilde{c}.H^T \equiv (c+e).H^T \tag{25}$$

where $H^T$ is the parity check matrix and $e$ is the error pattern belonging to the syndrome. The error pattern is expressed as:

$$e = \{q_1, q_2, \ldots, q_n\}, q_i \epsilon \{0,1\} \tag{26}$$

The error pattern may contain $s$ number of 1's ($s$ is equal to 1 for Hamming) which corresponds to the number of errors that are corrected in the code-word:

$$e_s = \{q_1, q_2, \ldots, q_n\} \mid num(q_i = 1) = s \tag{27}$$

To this end, every syndrome leads to exactly one error pattern with a single error. Therefore, there are $n$ unique patterns possible in $e_1$:

$$d_i^m = \{q_1, q_2, \ldots, q_n\} \iff (q_m = 1),$$
$$\forall (1 \leq i,j \leq n)\ d_1^i, d_1^j \in e_1, z \equiv (c+d_1^i).H^T \equiv (c+d_1^j).H^T \Rightarrow i=j \tag{28}$$

Meaning that each and every single error pattern would result in a unique syndrome. Furthermore, if the error pattern is all zeros, the code-word is correct, otherwise, decoding can be performed based on a syndrome table.

The same single error pattern corresponds to error patterns with 2 (duets) or 3 (triplets) errors. In order to distinguish between single error occurrence and higher number of errors, an extra parity bit is added (constructing a SECDED coding). This parity bit clarifies whether there was a single error in the code-word or two errors. If there is only one error, the decoder asserts the error based on the individual single error pattern:

$$\hat{c} = \tilde{c} + e \tag{29}$$

In which $\hat{c}$ is the corrected code-word. On the other hand, if two errors are detected, the error patterns can be expressed as $d_2^m \epsilon e_2$. However, there are several double error patterns that correspond to the same syndrome:

$$\exists (1 \leq i,j \leq n)\ d_2^i, d_2^j \in e_2, z \equiv (c+d_2^i).H^T \equiv (c+d_2^j).H^T\ \&\ i \neq j \tag{30}$$

Therefore, if there is no additional information, the code-word cannot be uniquely selected and the normal decoder would assert an error to the output.

On the other hand, in FaECC scheme, we consider these double-error pattern code-words and resolve which one should be considered to calculate the correct word. For this purpose, let us call the two nonzero bits in every $d_2^m$ "active bits". Under such conditions, these error patterns are orthogonal, meaning that for every $i,j$ that satisfies Eq. 27 for the same syndrome, we have:

$$\forall \{i,j, i \neq j\}, z \equiv (c+d_2^{i,j}).H^T \Rightarrow\ d_2^i.d_2^j = \vec{0} \tag{31}$$

Particularly, each of these code-words contain unique pairs of active bits; if bit $i$ and bit $j$ are active in code-word $x$, neither of them are active in any of the remaining code-words that satisfy Eq. 22 for the same syndrome as $x$. In other words, each and every specific bit in the code-word is active in at most one of the possible error-patterns. In order to identify the correct candidate error-pattern, there is a need to identify the location of one of the active bits. If both of the erroneous bits are soft errors, there is no way to find out which bits were erroneous. However, if one of the errors is a stuck-at-fault, meaning that it is possible to detect the location of the error, the correct value of both of the bits can be retrieved.

Table I. FaECC correction capability table.

| Type of error | SECDED | FaECC | DECTED |
|---|---|---|---|
| Number of encoded bits | $deg$+1 | $deg$+1 | $2deg$+1 |
| | Is the error corrected? | | |
| one soft or hard | yes | yes | yes |
| two hard | no | yes | yes |
| one soft and one hard | no | yes | yes |
| two soft | no | no | yes |

In order to find the location of one of the active bits, the erroneous code-word can be inverted and rewritten into the same line and read from it [Chen et al., 1984]. This inversion enables the decoder to detect any stuck-at-fault location and would assist in finding the correct code-word. At the next step, the code-word that was read the second time is compared to the code-word that was read at the first time and the location of the faulty bit(s) are derived. Eventually, active bits associated with the location(s) of faulty bits are fixed. Thus, the correct candidate code-word can be selected and the corrected word can be retrieved.

As explained above, the decoding scheme is capable of correcting a single error, two stuck-at-faults or a single stuck-at-fault and a single soft error. Table I compares the correction capability of SECDED, DECTED and FaECC. Where $deg$ is the degree of the Galois Field used to realize the coding scheme. If there are two soft errors, this scheme will not be able to correct it and will assert a fault as the output. Furthermore, although we used this scheme to correct errors in STT-MRAM memory arrays, it can be used to improve the yield of any type of memory array under the aforementioned conditions.

## 5.  CROSS LAYER SIMULATION FRAMEWORK

In order to analyze the reliability of STT-MRAM memory arrays, we developed a cross-layer simulation framework that captures the impact of various design parameters at different levels of abstraction (device, circuit, and architecture) on STT-MRAM memory array reliability. Fig. 7 shows the simulation framework and its different stages of analysis. The framework takes MTJ characteristics, memory specifications, and design constraints as inputs, and optimizes the memory array for the desired efficiency. We describe the simulation framework and its models at each level of abstraction in further detail next.

### 5.1 Device level

The simulation framework utilizes the device level model based on [Fong et al., 2012], which consists of a magnetization dynamics solver and a Non-Equilibrium Green's Function (NEGF) based electron transport solver [Danielewicz, 1984]. Initially, the NEGF solver is utilized to obtain $RA_{P,AP}$ vs. $T_{MgO}$ and $V_{MTJ}$. Next, the magnetization dynamics is obtained from the critical switching currents of the free layer, $J_C(AP➜P)$ and $J_C(P➜AP)$. The free layer is modeled as a monodomain ferromagnet. The magnetization of the monodomain ferromagnent is simulated by solving the Landau-Lifshitz-Gilbert equation, including the Slonczewski spin-torque term (LLGS) [Lee et. al., 2005].

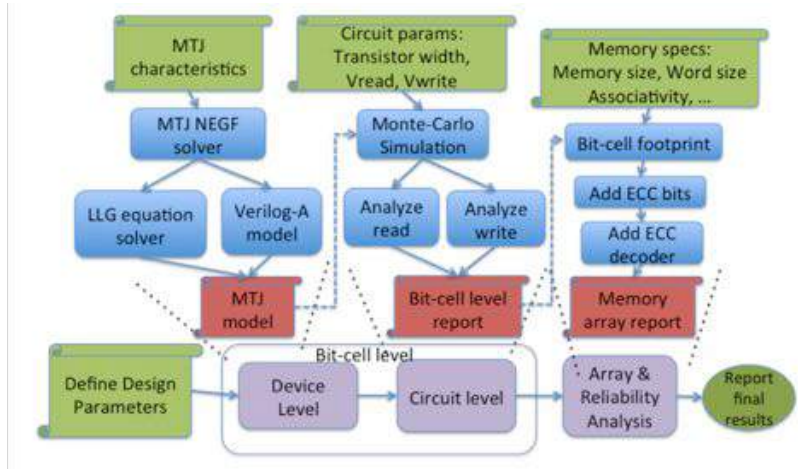

Fig. 7. Multilevel simulation framework.

$$\frac{d\hat{m}_{FL}}{dt} = \gamma\left(\vec{H}_{EFF}\times\hat{m}_{FL}\right) + \alpha\left(\hat{m}_{FL}\times\frac{d\hat{m}_{FL}}{dt}\right) + \gamma a_J(\theta)(\hat{m}_{FL}\times\hat{m}_{FL}\times\hat{m}_{PL}) \tag{32}$$

$$a_J(\theta) = \frac{\hbar J_{MTJ}}{2qM_St_{FL}}g(\theta) \tag{33}$$

$$g(\theta) = [-4 + \frac{(1+P)^3(3+cos\theta)}{4P^{1.5}}]^{-1} \tag{34}$$

where $\hat{m}_{FL}$ and $\hat{m}_{PL}$ are the unit magnetization vectors of the free layer (FL) and pinned layer (PL), respectively. Both FL and PL are considered to have the same $M_S$. $\gamma$ is the gyromagnetic ratio, $\alpha$ is the FL damping factor and $\vec{H}_{EFF}$ is the effective magnetic field. $q$ is the electronic charge and $J_{MTJ}$ is the current density through the MTJ and $P$ is the material-dependent spin polarization efficiency defined in [Slonczewski, 1996].

The characteristics of the MTJ are encapsulated in a Verilog-A model [Fong et al., 2012], which is used in HSPICE simulation [Hspice 2013.12]. Table II shows the device parameters assumed in this work. These and other bit-cell parameters were derived from [Fong et al., 2012], and the MTJ model was calibrated to experimental data published in literature [Yuasa et al., 2004].

### 5.2 Circuit level

The circuit level model of an STT-MRAM bit-cell consists of 32nm MOSFET models [Synopsys inc., 2014] and the MTJ Verilog-A model. HSPICE was used to simulate the circuit level behavior of the bit-cell. The load line method [Fong et al., 2012] was used to obtain probability of failure for different failure mechanisms. Fig. 8 illustrates the load line method. In this method, we consider variations in $t_{MgO}$ and cross-sectional area due to process variations. These variations affect $R_{MTJ}$. Variations in $R_{MTJ}$ affect the ability to write into the bit-cell (write failures), the ability to correctly sense $R_{MTJ}$ of the bit cell (decision failures), and the ability of the MTJ to retain its configuration when the bit-cell is being read (disturb failures). In order to determine write failures, it is considered that the MTJ cross-sectional area has a Gaussian distribution. For each MTJ cross-sectional area, the critical current density ($J_C$) is determined. At the next step, the transistor $I_D$-$V_{DS}$ (obtained using

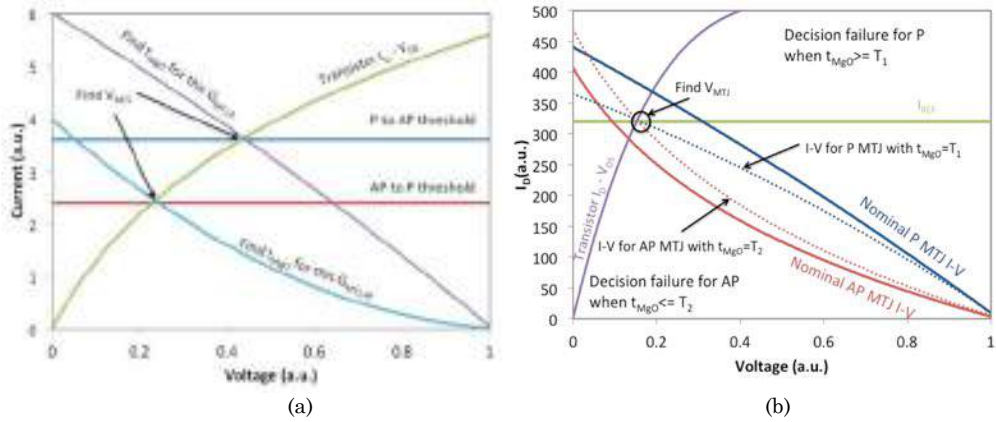

(a)

(b)

Fig. 8. Load line method illustration for (a) write and read-disturb failures, (b) read decision failure.

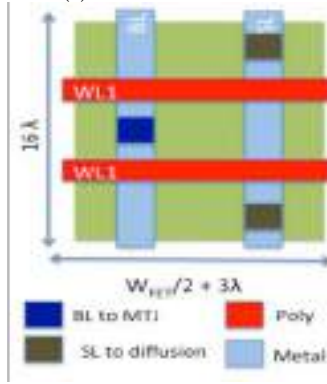

Fig. 9. Two-finger layout of STT-MRAM bit-cell.

Monte Carlo simulations in HSPICE) we find the voltage across the MTJ ($V_{MTJ}$) from the DC load line analysis as shown in Fig. 8 (a). Eventually, the maximum $R_{MTJ}$ (and the corresponding maximum $t_{MgO}$) that allows successful write in the MTJ is calculated. Hence, any bit-cell having an MTJ with the same area but a thicker $t_{MgO}$ will not be written in the targeted write time. Therefore, the bit-cell fails write operation. A similar analysis is performed for read disturb failures. However, in read disturb failures, the bit-cells with thinner $t_{MgO}$ are considered to fail.

Decision failures occur when the sense amplifier outputs $H$ for a bit-cell in P configuration ($R_L$) and $L$ for a bit-cell in AP configuration ($R_H$). The probability that a functioning sense amplifier senses the bit-cell configuration incorrectly is called the read decision failure. The reference current ($I_{REF}$) needs to be chosen to minimize this probability. For a bit-cell with an MTJ of a particular cross-sectional area, a certain $t_{MgO}$ will result in the bit-cell current to be $I_{REF}$. If the MTJ is in AP (P), a thinner (thicker) $t_{MgO}$ will result in a smaller (larger) $R_{MTJ}$ and a bit-cell current higher (lower) than $I_{REF}$. Fig. 8 (b) illustrates the method used to determine the decision failures for each $I_{REF}$. The optimum read reference current is the reference current that minimizes the read probability of failure. In our analysis, we perform a linear search between the nominal read currents of $P$ and $AP$ configurations to obtain the optimum reference current.

The variations in the MTJ considered were variations in the cross-sectional area and the oxide thickness. Both were considered to have normal distribution with 2% variance. Moreover, in order to capture the variations in the access transistor, 1e4 Monte-Carlo simulations were performed and the aforementioned methods were used to obtain the probability of failure for different failure mechanisms.

## 5.3 Array level

At the array level, CACTI [Muralimanohar et al., 2009] was modified to include the access time and energy model of the STT-MRAM. We considered a two-finger layout of the bit-cell as explained in [Gupta et. al., 2012]. The layout of the bit-cell is illustrated in Fig. 9. Additionally, the number of encoding bits was added to the CACTI model to capture the impact of the extra bit-cells on the memory efficiency.

In order to analyze the impact of ECC, we implemented the ECC codecs. The encoders for the Hamming code and the proposed FaECC were identical. However, the Hamming decoder and the FaECC decoder were different. For the Hamming decoder, our implementation was based on [Opencores, 2015]. For the FaECC decoder, an RTL HDL description was developed using the lookup table method [Howell et al., 1977]. Synopsys Design Compiler [Design Compiler, 2011] was then used to implement the decoders in the 32 nm Technology node. Table III shows the characteristics of the decoders.

In order to calculate the efficiency of the memory at the array level, for each write operation, it was considered that the bits were encoded and written into the memory; therefore, the overheads associated with encoding were calculated towards the total efficiency of the memory. However, for the read operation, the results were obtained based on the weighted average number of each of the three possible scenarios: 1) read operation and error detection, 2) read operation and single error correction, 3) read operation and double error correction using reread, rewrite and FaECC decoder.

## 6. RESULTS AND DISCUSSION

We designed a 1 MB STT-MRAM cache to evaluate the proposed ECC techniques. Table IV presents the characteristics of the cache. In order to capture the impact of process variations on the MTJ, the volume of the free layer was considered to have a variation with a standard deviation of 2% of the nominal value. The same variation level was considered for the cross section of the MTJ. Also, in order to analyze the impact of variations on the access transistor, as explained earlier, the load line method [Fong et al., 2012] was used to obtain read and write failures.

Initially, we investigated the read operation and analyzed the impact of different parameters on read operation reliability. Fig. 10 (a) illustrates the probability of read failure *vs.* the access transistor width. The read decision failure probability increases slightly with an increase in the transistor width. This is due to the degradation in the bit-cell TMR with higher transistor widths. Furthermore, the probability of failure is slightly higher for $V_{read}$=200mV; however, the difference is smaller than an order of magnitude. For read disturb failures, our results show that this type of failures are negligibly small for our design. Therefore, the read decision failures dominate the probability of read failure. This makes the probability of read failure virtually independent of $E_B$.

Next, we investigated write operation reliability. Fig. 10 (b) illustrates the probability of write failure *vs.* access transistor size for two different write pulse widths. The $V_{dd}$ was considered to be 1V. As observed, the probability of write failure decreases with an increase in the width of the access transistor. This is due to the increase in the write current for higher transistor widths. Due to process variations, some of the bit-cells have higher than nominal critical current; by increasing the write current, these bit-cells are successfully written. Therefore, the probability of

Table II. Parameters for MTJ

| | |
|---|---|
| Magnetization Orientation | Perpendicular |
| Nominal Free Layer volume | 64nm x 64nm x 1nm |
| Oxide thickness | 1nm |
| PMA Anisotropy Energy Barrier | 50kBT-70kBT |
| Gyromagnetic Factor, γ | 17.6 GHz/Oe |
| Saturation Magnetization, MS | 850 emu/cm3 |
| Damping factor, α | 0.028 |
| Temperature | 300°K |

Table III. Decoder synthesis results

| Decoder type | SECDED (128,137,1) | FaECC (128,137,*) | DECTED (128,145,2) |
|---|---|---|---|
| Area (um2) | 5433 | 114762 | 106700 |
| Delay (ns) | 1.45 | 10.47 | 3.62 |
| Dynamic power (mW) | 0.123 | 13.58 | 1.09 |
| Leakage power (mW) | 0.142 | 3.5 | 2.23 |

*1 soft error or 1 soft and 1 hard error or 2 hard errors

write failure decreases with an increase in the access transistor size. This trend is observed for both of the write pulse widths shown in Fig. 10 (b). However, the probability of failure is larger for the 6 ns pulse width compared to the 8 ns pulse width. This is due to the inverse relationship between the critical current of the MTJ and write pulse width – the critical current is smaller for 8 ns pulse width compared to 6 ns. Therefore, for a given nominal transistor width (which results in a given write current), the number of bit-cells with currents less than the critical current of the bit-cell is smaller for 8 ns compared to 6 ns.

Let us consider the relationship between $E_B$ and the probability of write error. Fig. 11 illustrates the probability of write error with respect to access transistor width for different values of $E_B$ for a fixed pulse width of 8 ns. As observed, for a given transistor width, the probability of error is higher for higher $E_B$. This relation stems from increased the critical current of the bit-cell with higher $E_B$.

Once the design space was explored and the bit-cells were characterized, we designed caches optimized for different design metrics. For this purpose, we considered the target yield to be 99.9% and the run-time reliability to be 1 FIT. As observed in Fig. 10 (a), the read probability of failure is of the order of 1e-6 and does not change substantially with change in the transistor width. Furthermore, in order to define the write probability of failure, the nominal write pulse width of 8 ns is used. As observed in Fig. 10 (b), the write failure increases drastically with a decrease in the access transistor width. Therefore, if the read and write probabilities of failure are considered jointly, the probability of bit-cell failure cannot be made lower than ~1e-6 by adjusting the transistor size and/or the read voltage. Therefore, it is not possible to meet the reliability target without applying ECC. This result matches the results in [Xu et al, 2009; Del Bel et al., 2014; Pajouhi et al., 2015; Kwon et al., 2015].

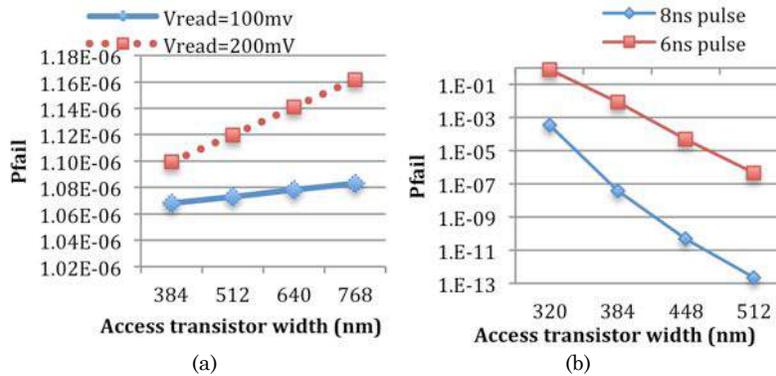

(a)                                         (b)

Fig. 10. Bit-cell reliability analysis: (a) The probability of read error for two different read voltages, (b) probability of write failure vs. access transistor width.

Next, we considered a cache with ECC. In order to have a fair comparison with respect to different ECC schemes, we considered a 128-bit ECC for SECDED and FaECC and DECTED.

Fig. 12 compares the energy, the area and the read/write latency for caches with different ECC configurations when the caches were optimized for minimum area. As observed, the area for a cache with FaECC is 20% and 13% less than a cache with SECDED and DECTED, respectively. Notably, the access transistor width is smaller for FaECC compared to SECDED. This reduced transistor width stems from higher coverage for read and write errors for FaECC compared to SECDED. Furthermore, since SECDED and DECTED are used for yield enhancement as well as for enhancing run-time reliability, it may not be easy to reduce $E_B$ (run-time errors may increase). On the other hand, if FaECC is used, there exists an opportunity to optimize $E_B$, while still maintaining good coverage for run-time errors.

Next, we optimized the cache for minimum energy consumption. In order to have a fair comparison between the three cache configurations, we considered the mean energy consumption of every read operation. Specifically, the energy associated with the error detection unit is considered for each and every read operation. However, for SECDED and DECTED, the decoder energy is considered only when an error is detected. Further, for FaECC, if a single error is detected, the decoding procedure would involve correcting a single error; thus, it would not include the additional write and read step. On the other hand, if two errors are detected, the decoding would involve extra write and read operations as well as the use of the additional decoding step. Therefore, the energy associated with each of these two conditions is added to the total energy based on the number of times each condition is applicable. Fig. 13 compares the read/write energy and the area of the cache after energy optimization is performed. As observed, the read energy of the cache with FaECC is 8% less than SECDED and 4% less than DECTED. However, the write energy of FaECC is 21% and 11% smaller than that of caches with SECDED and DECTED, respectively. Also, as observed in Fig. 13, the area of the cache with FaECC is 20% and 11% less than the caches with SECDED and DECTED, respectively.

Table IV. Cache characteristics

| Block size (bytes) | 16 |
|---|---|
| Associativity | 1 |
| Read/Write port(s) | 1 |
| Technology | 32 nm |
| Cache model | Uniform Cache Architecture (UCA) |

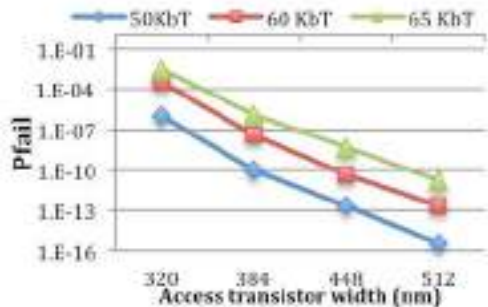

Fig. 11. Write error probability *vs.* access transistor width for different $E_B$.

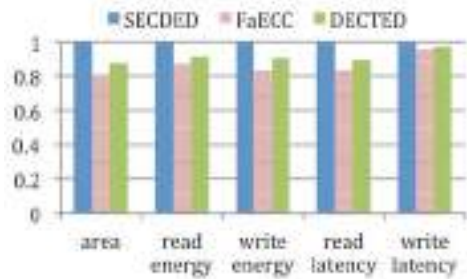

Fig. 12. Area, energy, and latency for caches with different ECC schemes.

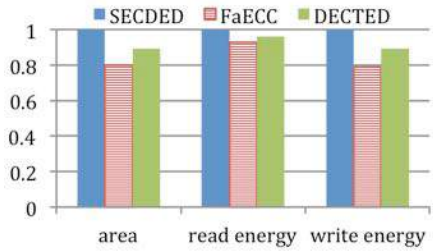

Fig. 13. Cache optimized for energy.

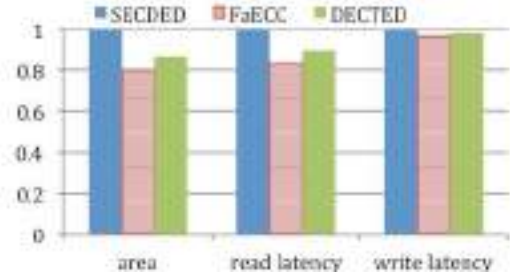

Fig. 14. Cache optimized for performance.

We also optimized the cache for improved write performance. In order to have a fair comparison, we compared the mean delay of the three different ECC schemes. For this purpose, the write latency was calculated based on the latency required for a successful write operation as well as the latency for calculating the encoding bits. In order to obtain the mean read latency, similar to calculating the energy, we considered the weighted average of the delays of different ECC schemes based on the number of times they are invoked. For all ECC schemes, the error detection delay is included in every read operation. However, for SECDED and DECTED, the decoder delay is considered only when an error is to be corrected. For FaECC, as observed in Fig. 6, the data detection is performed for every data read and the correction unit is used only if there is an error.

If there were only a single error, the FaECC decoder would have the same delay as a SECDED decoder. However, the FaECC scheme differs from SECDED when two errors are detected. Additionally, this occurs only if the data value written into the hard error location is different from the data read from that location: if a bit-cell with a hard error of "0" is storing "0" (the same value) it will be read without any error. When FaECC is activated, the read latency would be dominated by a write and a read operation. This is due to parallel estimation of the candidate code-words and the additional write and read procedure. Note, that the hard error locations are required only at the end of the correction procedure. Therefore, the worst-case delay associated with FaECC is longer than that of SECDED or DECTED.

On the other hand, in STT-MRAMs, due to the long latency associated with STT switching, the write pulse width dominates the write latency. Therefore, for write performance optimization, the write pulse width should be reduced. In order to have a fair comparison, we reduced the write pulse width of all three ECC configurations to 6 ns and optimized the cache for performance. It can be observed from Fig. 10 (b) that if the write pulse duration is equal to 6 ns instead of 8 ns, the probability of write failure increases for a fixed access transistor width. In order to compensate for this increased probability of failure, the access transistor can be upsized to ensure complete STT switching. However, upsizing the access transistor negatively impacts the read performance. Therefore, there is a trade-off between the read performance and the write performance. Fig. 14 depicts the area and read/write latency for a cache with different ECC configurations. As observed, the read latency of FaECC is 16% less than that of the cache with SECDED and 11% less than that of the cache with DECTED. Furthermore, the area is 19% and 14% less than that of SECDED and DECTED, respectively.

## 7.  CONLUSION

In this paper, we analyzed the impact of process variations on the run-time reliability of STT-MRAM memory arrays. Furthermore, we analyzed the efficacy of ECC in relaxing $E_B$ requirement of the MTJ under process variations. We also analyzed the efficacy of ECC on yield enhancement and run-time reliability. Our results showed that if SECDED is used for yield enhancement besides run-time reliability, it may be difficult to have a more relaxed value of $E_B$ (better write current). Thus, we proposed using FaECC in which permanent faults are masked while maintaining its correction capability for soft errors. In order to analyze the efficacy of FaECC, we developed a simulation framework that considers different levels of design abstraction.  Using the simulation framework, we performed a case study of a 1 MB cache in 32 nm Technology node. We showed that in our proposed scheme, the area of the memory array is reduced up to 20% compared to a cache with SECDED and up to 13% compared to a cache with DECTED, at iso-reliability. Furthermore, the write energy can be reduced up to 21% and 11% compared to caches with SECDED and DECTED correction capabilities, respectively.